\documentclass{article}
\usepackage{spconf,graphicx}
\usepackage{amsmath,amssymb,amsfonts}
\usepackage{algorithmic}
\graphicspath{{./Images/}}
\usepackage{textcomp}
\usepackage{xcolor}
\usepackage{subfig}
\usepackage{caption}
\usepackage{url}
\usepackage{tikz}
\usepackage{hyperref}

\def\Figu#1{{Figure~\ref{#1}}}

\def\Tabl#1{{Table~\ref{#1}}}
\def\comment#1{}

\title{Image Segmentation for Improved Lossless Screen Content Compression}
\twoauthors
{Shabhrish Reddy Uddehal, Tilo Strutz \sthanks{This research work is funded by Deutsche Forschungs\-ge\-mein\-schaft (DFG, German Research Foundation) - 438221930.}}
{Department of Electrical Engineering \\
and Computer Science \\
Coburg University, Coburg \\
Germany}
{Hannah Och, Andr\'{e} Kaup}
{Multimedia Communications   \\
and Signal Processing \\
Friedrich-Alexander University, Erlangen\\
Germany}

\begin{document}
%

\newcommand\copyrighttext{%
	\footnotesize \textcopyright 2023 IEEE. Personal use of this material is permitted. Permission from IEEE must be obtained for all other uses, in any current or future media, including reprinting/republishing this material for advertising or promotional purposes, creating new collective works, for resale or redistribution to servers or lists, or reuse of any copyrighted component of this work in other works. DOI: \url{https://ieeexplore.ieee.org/document/10094983}. 
}
\newcommand\copyrightnoticeOwn{%
	\begin{tikzpicture}[remember picture,overlay]
		\node[anchor=north,yshift=-10pt] at (current page.north) {\fbox{\parbox{\dimexpr\textwidth-\fboxsep-\fboxrule\relax}{\copyrighttext}}};
	\end{tikzpicture}%
	\vspace{-8mm}
}

\maketitle
\copyrightnoticeOwn
\begin{abstract}
In recent years, it has been found that screen content images (SCI) can be effectively compressed based on appropriate probability modelling and suitable entropy coding methods such as arithmetic coding. The key objective is determining the best probability distribution for each pixel position. This strategy works particularly well for images with synthetic (textual) content. However, usually screen content images not only consist of synthetic but also pictorial (natural) regions. These images require diverse models of probability distributions to be optimally compressed. One way to achieve this goal is to separate synthetic and natural regions. This paper proposes a segmentation method that identifies natural regions enabling better adaptive treatment. It supplements a compression method known as Soft Context Formation (SCF) and operates as a pre-processing step.
If at least one natural segment is found within the SCI, it is split into two sub-images (natural and synthetic parts) and the process of modelling and coding is performed separately for both. For SCIs with natural regions, the proposed method
achieves a bit-rate reduction of up to 11.6\% and 1.52\% with respect to HEVC and the previous version of the SCF.

\end{abstract}
%

\section{Introduction}
\label{sec:intro}

The increasing number of connected digital devices demands new technologies for transmitting screen content. Typically, screen content images refer to images that are generated when taking screenshots or during desktop sharing of computer displays. The 1:1 transfer of screen content from a computer to a remote display is one of the main applications for processing screen content data. Computer screens usually show text and graphics (synthetic) in combination with pictorial elements (natural). The synthetic parts are mainly characterised by structures such as text, icons, window frames, and buttons. They have a limited number of colours and many repetitive patterns. In comparison, natural parts contain very realistic content rendered by a computer or captured by a camera. These parts have a very high number of unique colours and few to none repetitive patterns.

The standardization of High Efficiency Video Coding (HEVC) introduced measures to improve the compression performance for image sequences containing still and moving graphics, text or camera-captured content (\cite{Sullivan12, Xu16a}).
The extension to screen content coding (HEVC-SCC) incorporates coding tools such as intra block copy \cite{Xu16d}, adaptive colour transform \cite{Zhang15}, and palette mode \cite{Pu16}. With further modifications, the above mentioned tools are integrated into the VVC standard \cite{Ngu21}. With respect to lossless coding of screen content, both HEVC-SCC and VVC contain similar coding tools, therefore have very similar compression performance.

One possible approach for efficient and content adaptive coding of screen content images is based on segmentation. The idea is to decompose a compound image into multiple layers or blocks and then process them separately. For block based segmentation, the image is typically sliced into 8 x 8 or 16 x 16 non-overlapping blocks which are classified into textual or pictorial blocks based on statistical properties such as standard deviation \cite{Eben20}, using a colour threshold (\cite{Lin05, deQ00}), histograms and gradients of the blocks \cite{Din06}.

In \cite{Mina15a}, the least absolute deviation approach is used to segment the foreground layer containing text and the background layer containing smoothly varying regions. A sparse decomposition framework is proposed for background/foreground segmentation in \cite{Mina15b}. Local image activity measure algorithm is proposed in \cite{Yang15}, 
to measure the local pixel variations for the segmentation of textual and pictorial regions. In \cite{Yang12}, an approach based on text dictionary learning has been proposed for the compression of text regions, the separation of synthetic and natural regions is performed in the wavelet domain.
A fully convolutional neural network is used for the segmentation of curved text from images in \cite{Xu19}.

The most effective method for compressing synthetic content is Soft Context Formation (SCF) \cite{Str19}. Based on adaptively generated probability distributions, it attempts to provide an arithmetic coder with optimal information. However, as soon as the image to be compressed contains regions with different statistical properties, this method becomes less effective. The probability modelling can be improved if it is informed about these regions.  

This paper proposes a new method that identifies rectangular segments with natural content. It first classifies small image blocks into natural and synthetic blocks based on the number of unique colours. Then the natural regions are extracted by their bounding box coordinates. If at least one segment is found within the SCI, the natural and the synthetic regions are coded separately.

\section{Soft context formation}	
\label{sec_SCF}

SCF \cite{Str19} is the most recent and successful approach for lossless coding of SCI. For each single pixel position, a probability distribution is derived based on accumulated statistics of already encoded pixels and is fed into a fully adaptive arithmetic coder \cite{Str16b}. While encoding the pixels, the image statistics are learned in the form of a list of already encountered patterns (called pattern list) and a global colour palette containing all previously seen colours and their counts. A pattern comprises a template with the colours of the closest six causal neighbours with respect to the current pixel position.
The pixels are coded from the top right corner of the image to the bottom left corner (raster-scan). Based on the statistics accumulated by already encoded pixels, the probability distribution is estimated for each upcoming pixel. 

\begin{figure}
\hfil\usetikzlibrary{shapes.geometric}
\usetikzlibrary{positioning}
%
%

\tikzstyle{block} = [draw, fill=white, rectangle, minimum height=1.5em, minimum width=5em, align=center]
\tikzstyle{decision} = [draw, fill=white,diamond, aspect=2,minimum height=2em, minimum width=8em,align=center, inner sep=0pt, outer sep=0pt]
\tikzstyle{input} = [coordinate]
\tikzstyle{output} = [coordinate]
\tikzstyle{pinstyle} = [pin edge={to-,thin,black}]
\tikzstyle{pinstyle2} = [pin edge={-to,thin,black}]
\tikzset{font=\scriptsize}

\begin{tikzpicture}
	%
	%
	\noindent
    \node [decision, pin={[pinstyle]above:Get current pixel $X$}, node distance=1.5cm] (stage1decision){$X$ can be\\ coded in Stage 1};
		\node [decision,below of=stage1decision, node distance=1.5cm] (stage2decision){$X$ is in\\ colour palette};
		\node [block, right of=stage1decision, node distance=3.5cm] (stage1){Code $X$ in Stage 1};
		\node [block, right of=stage2decision, node distance=3.5cm] (stage2){Code $X$ in Stage 2};
		\node [block, below of=stage2, node distance=1.25cm] (stage3){Code $X$ in Stage 3};
		\node [block, below right=0.25cm and -0.25cm of stage3,pin={[pinstyle2]below:Goto next pixel}] (update){Update histograms\\and pattern lists};
		
		\draw  [->] (stage1decision) -- node [name=Yes1, midway, above] {Yes} (stage1);
		\draw  [->] (stage2decision) -- node [name=Yes2, midway, above] {Yes} (stage2);
		\draw  [->] (stage1decision) -- node [name=No1, midway, left] {No} (stage2decision);
		\draw [->] (stage2decision) |- node [name=No2, near start, left] {No} (stage3);
		\draw [->] (stage1) -| (update);
		\draw [->] (stage2) -| (update);
		\draw [->] (stage3) -| (update);
\end{tikzpicture}


		%
		%
		%
 ~\\[-3em]
\caption{\label{block_diagram} Block diagram of SCF coding for single pixel \cite{Och21}}
\end{figure}
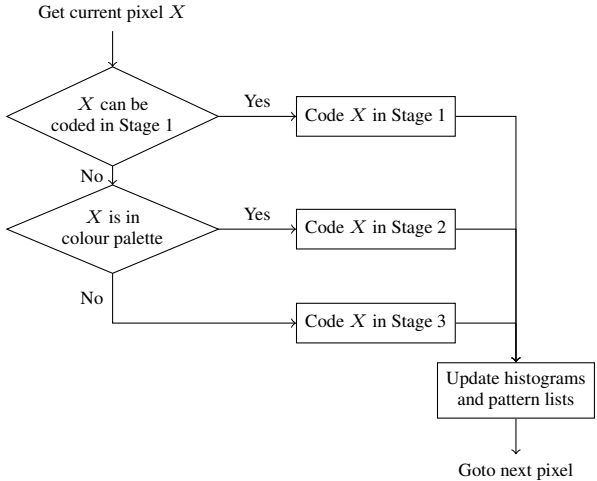

Let us assume that \emph{X} is the current pixel (RGB triplet) to be encoded. It can be coded in one of three stages as shown in \Figu{block_diagram}. 
If the colour of \emph{X} has already appeared in combination with a pattern that is similar to its current one, \emph{X} is coded in Stage 1. If there is no similar pattern in the pattern list or the colour is not associated with any similar pattern, an escape symbol is transmitted to initiate Stage 2 coding. If the colour of \emph{X} has already been seen in the image, it can be encoded in Stage 2 based on the global colour palette. In case that \emph{X} is a completely new colour, another escape symbol has to signal the transition to Stage 3. This stage tries to predict the three colour components separately and maintains probability distributions for prediction errors \cite{Och21}. Stage 1 is the preferred coding stage because of its coding efficiency and retention of lower bit-rates in comparison to Stages 2 and Stage 3.

\section{Proposed method}	
\label{sec_segmentation}

As already discussed in previous sections, the mix of synthetic and natural content can have adverse effects on the compression efficiency because it makes the estimation of proper distributions difficult.
To handle this issue, we propose a segmentation algorithm to automatically detect natural regions, such that the synthetic background is encoded separately from natural segments. This approach offers more flexibility over the learned statistics. \Figu{fig_SegmentationSteps} depicts the general stages of the proposed segmentation process which can be clustered in three intermediate steps, namely block classification (\Figu{fig:sfig2}), segmentation refinement (\Figu{fig:sfig3} - \Figu{fig:closeUp}) and processing (\Figu{fig:sfig6}, \Figu{fig:sfig7}).

\subsection{Block classification}	
	

\begin{figure}[t!]
\centering
  \subfloat[Original image]
	  {\frame{\includegraphics[height=3.5cm]{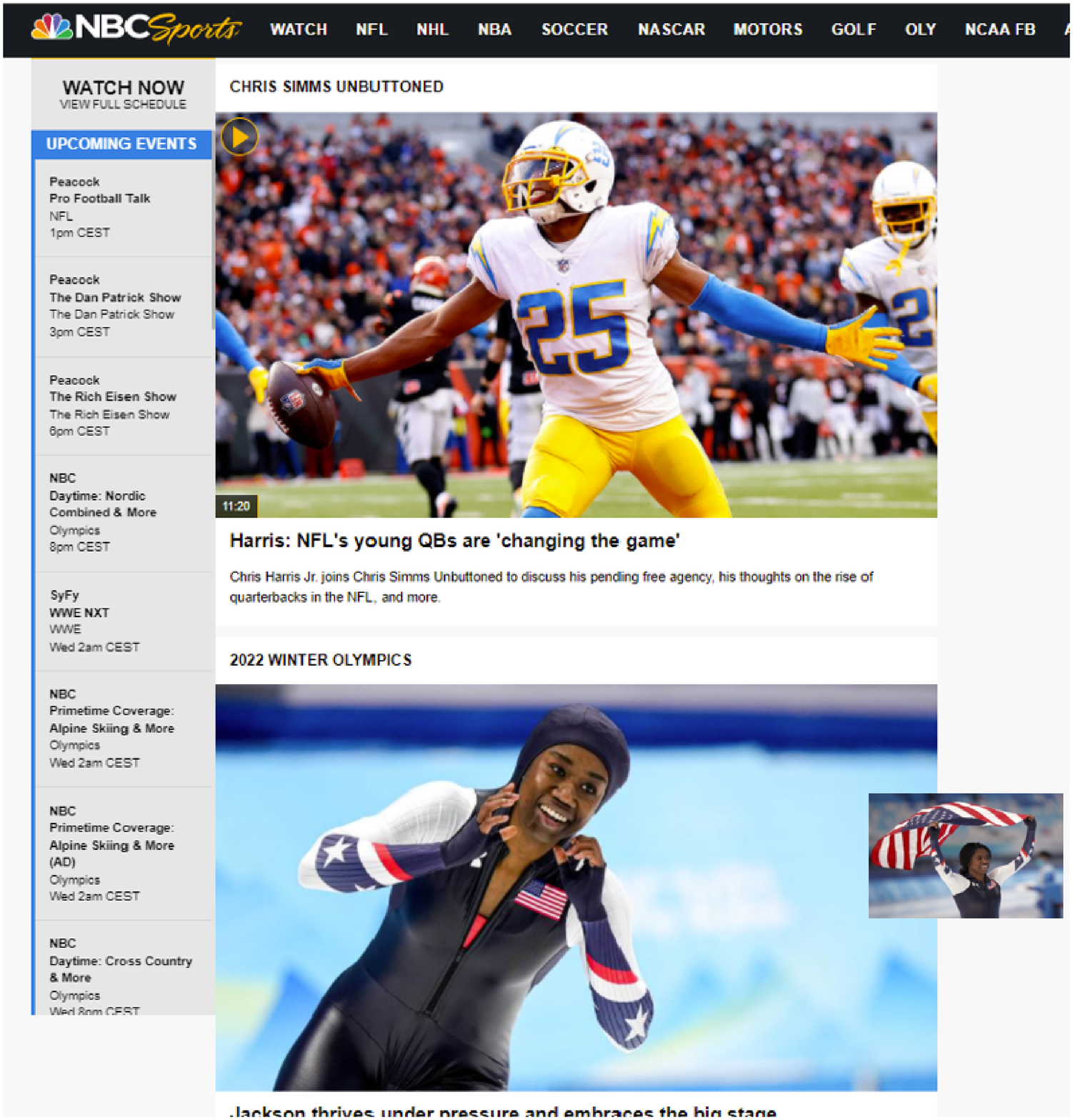}}\label{fig:sfig1}}
	\hfil
  \subfloat[Block classification of image]
	  {\frame{\includegraphics[height=3.5cm]{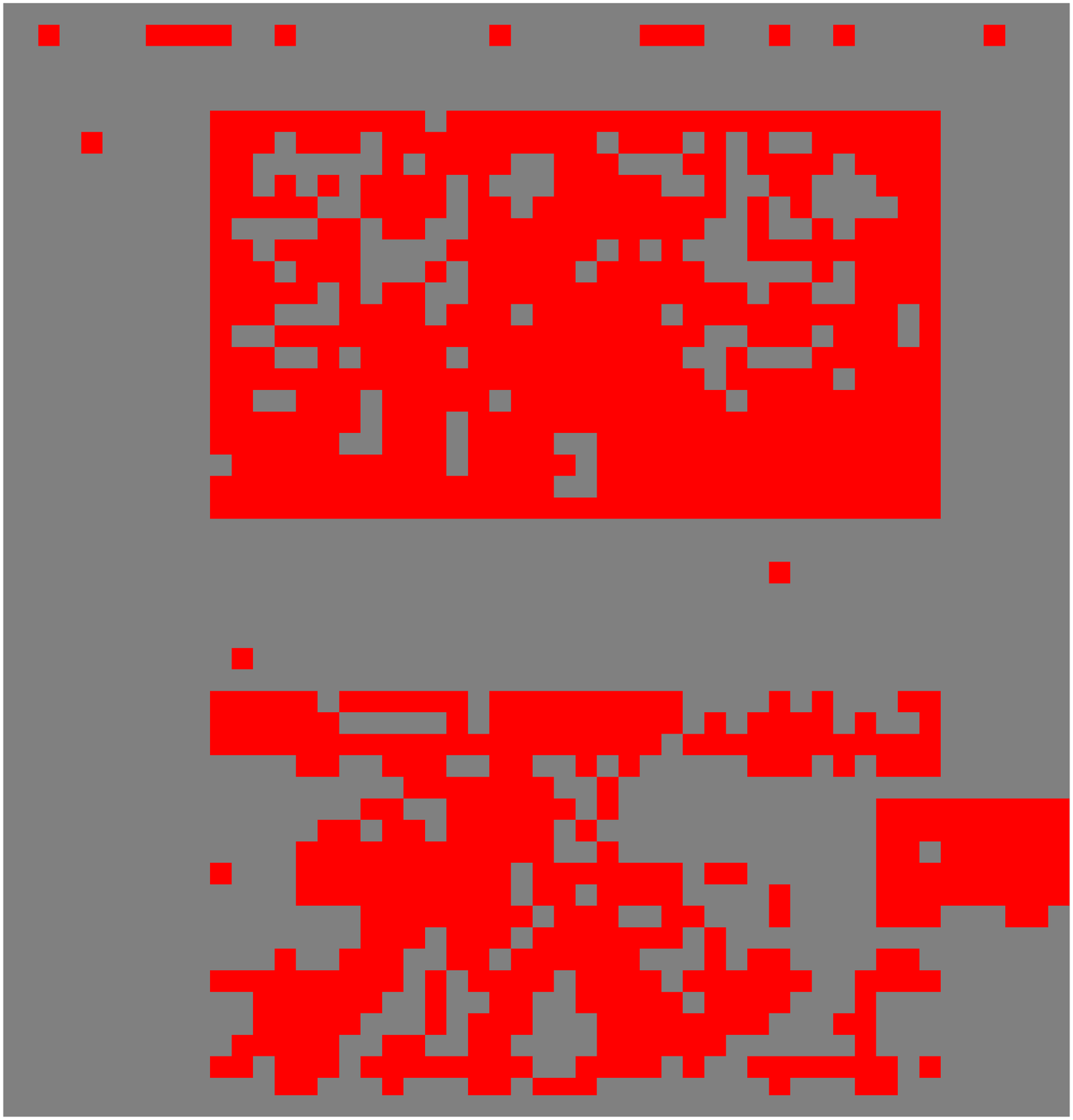}}\label{fig:sfig2}}

	\subfloat[Initial bounding boxes]
		{\frame{\includegraphics[height=3.5cm]{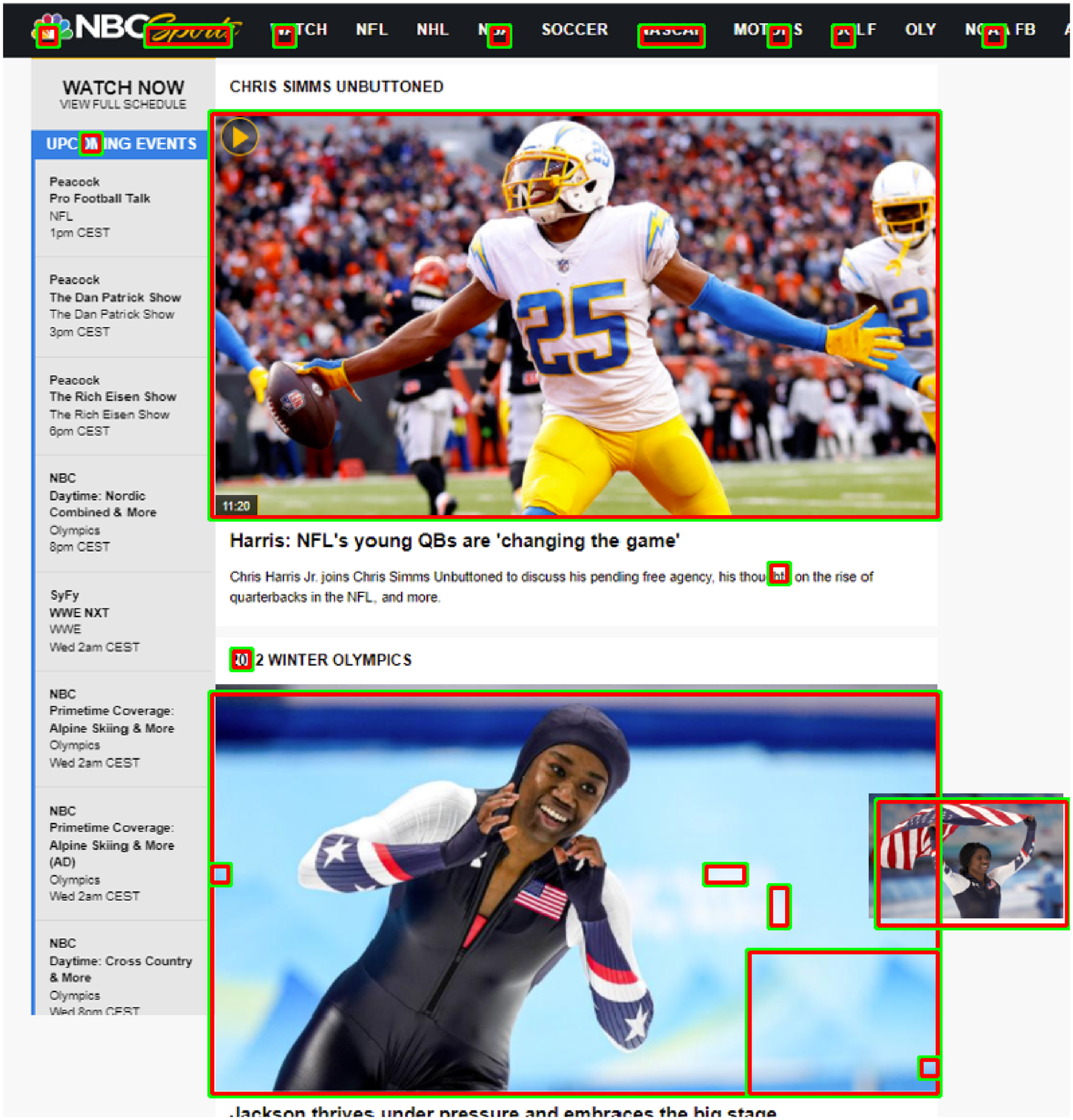}}\label{fig:sfig3}}
	\hfil
  \subfloat[After area threshold]
		{\frame{\includegraphics[height=3.5cm]{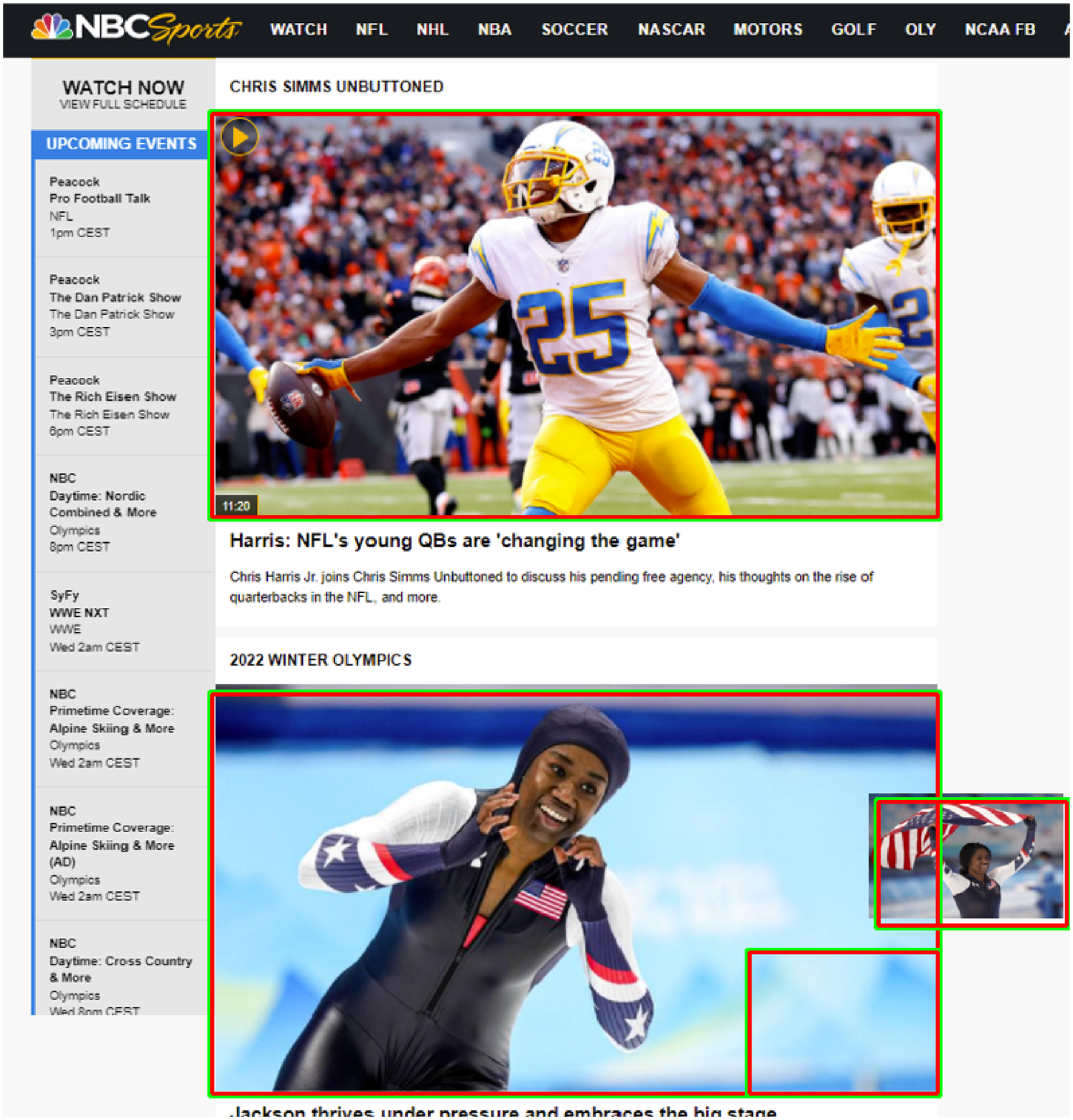}}\label{fig:sfig4}}

	 \subfloat[After refinement]
		{\frame{\includegraphics[height=3.5cm]{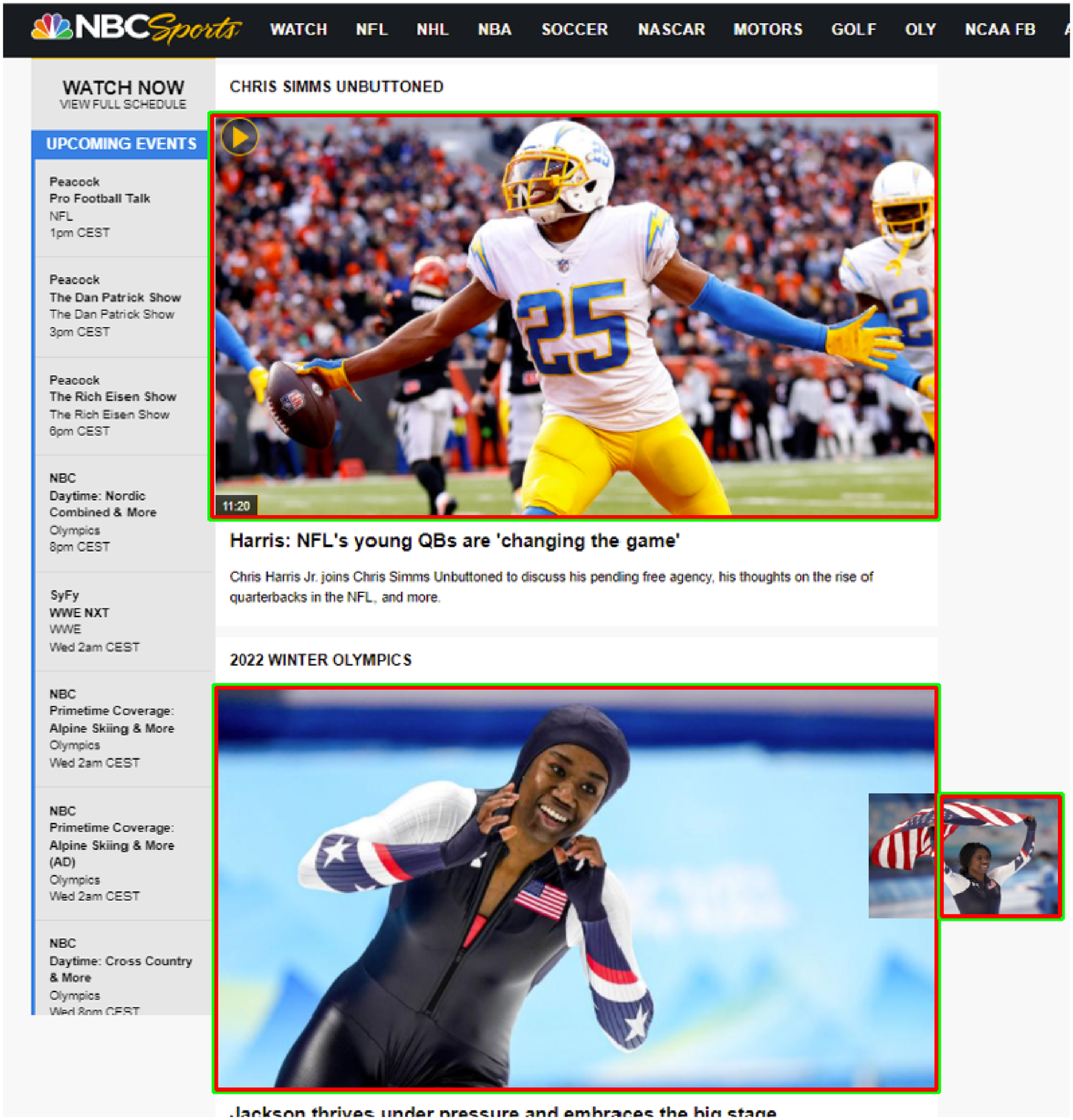}}\label{fig:sfig5}}	
		\hfil
	 \subfloat[Enlarged parts of \ref{fig:sfig4}, \ref{fig:sfig5}]
		{\includegraphics[height=3.5cm]{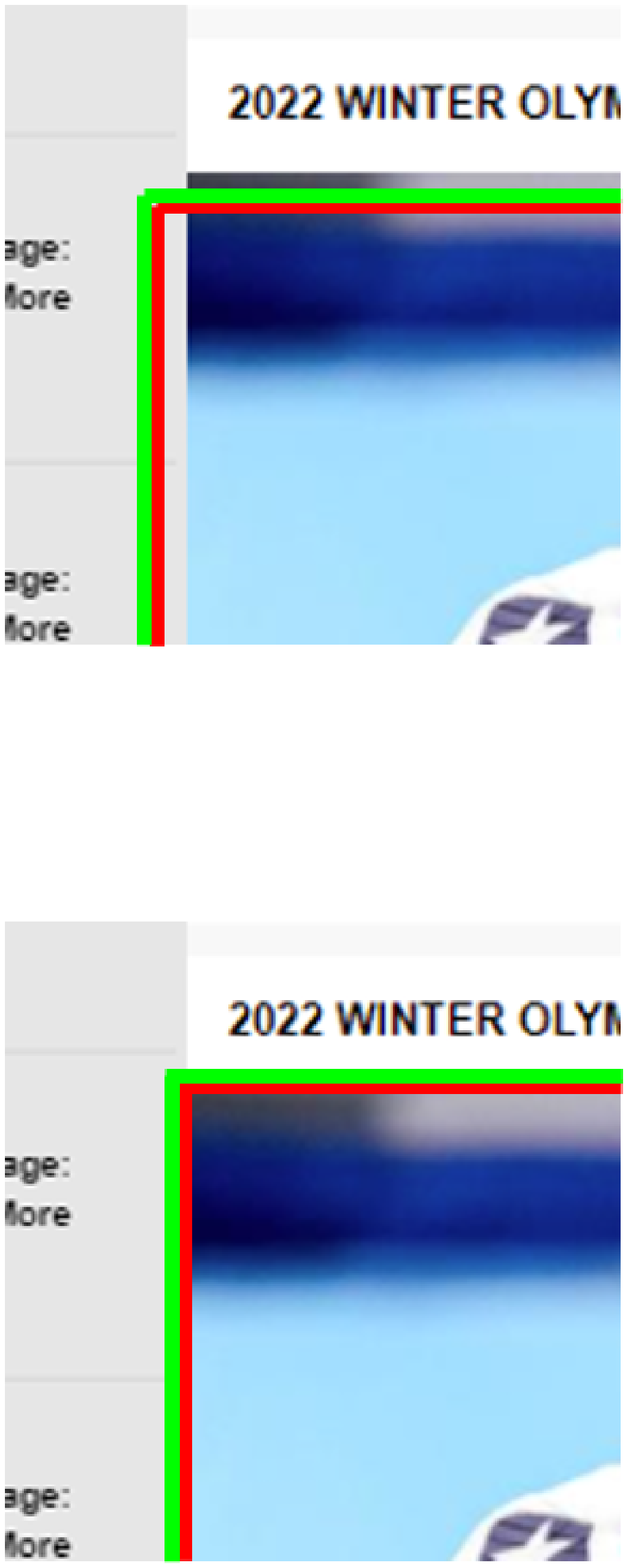}
		\hfil\includegraphics[height=3.5cm]{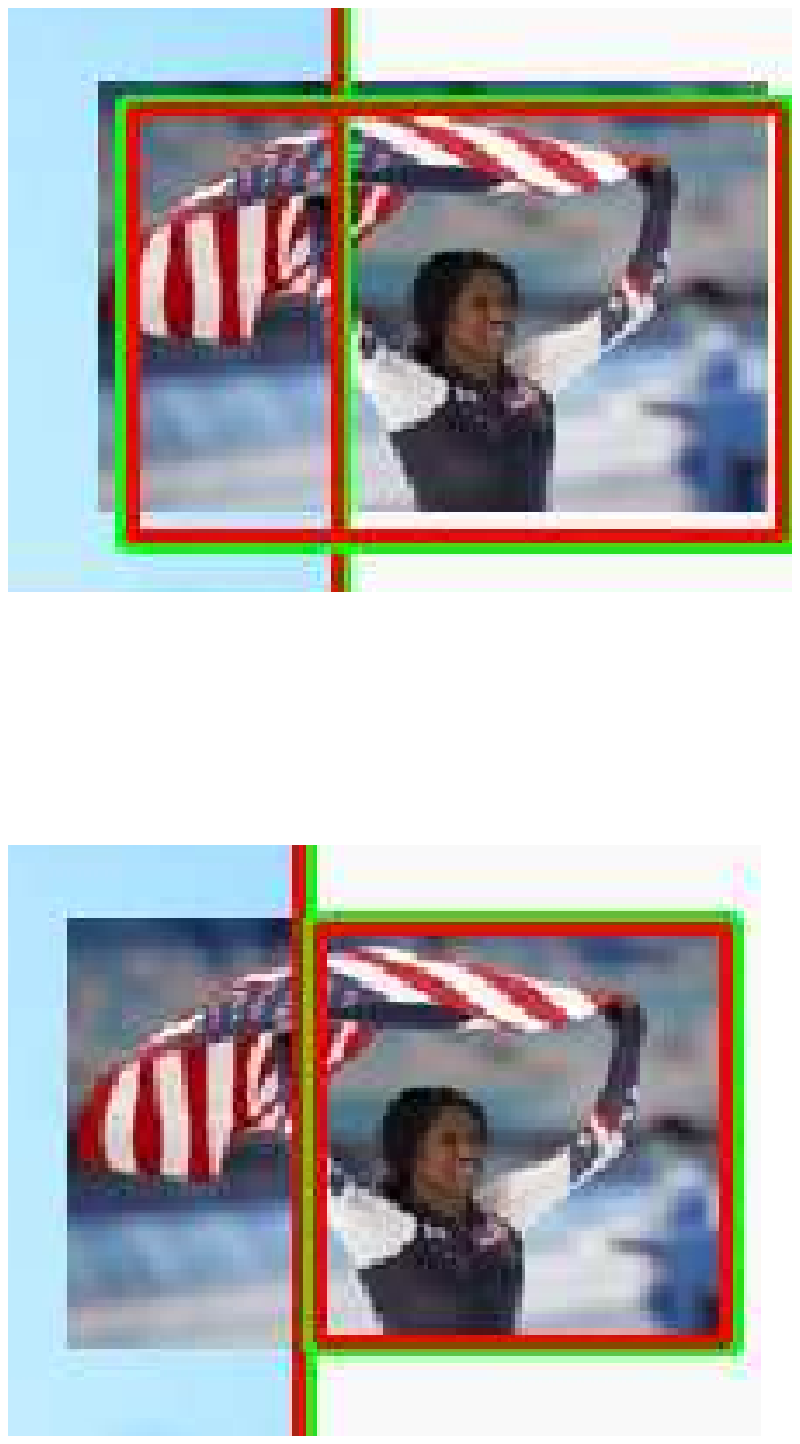}\label{fig:closeUp}}

  \subfloat[Synthetic background]
		{\frame{\includegraphics[height=3.5cm]{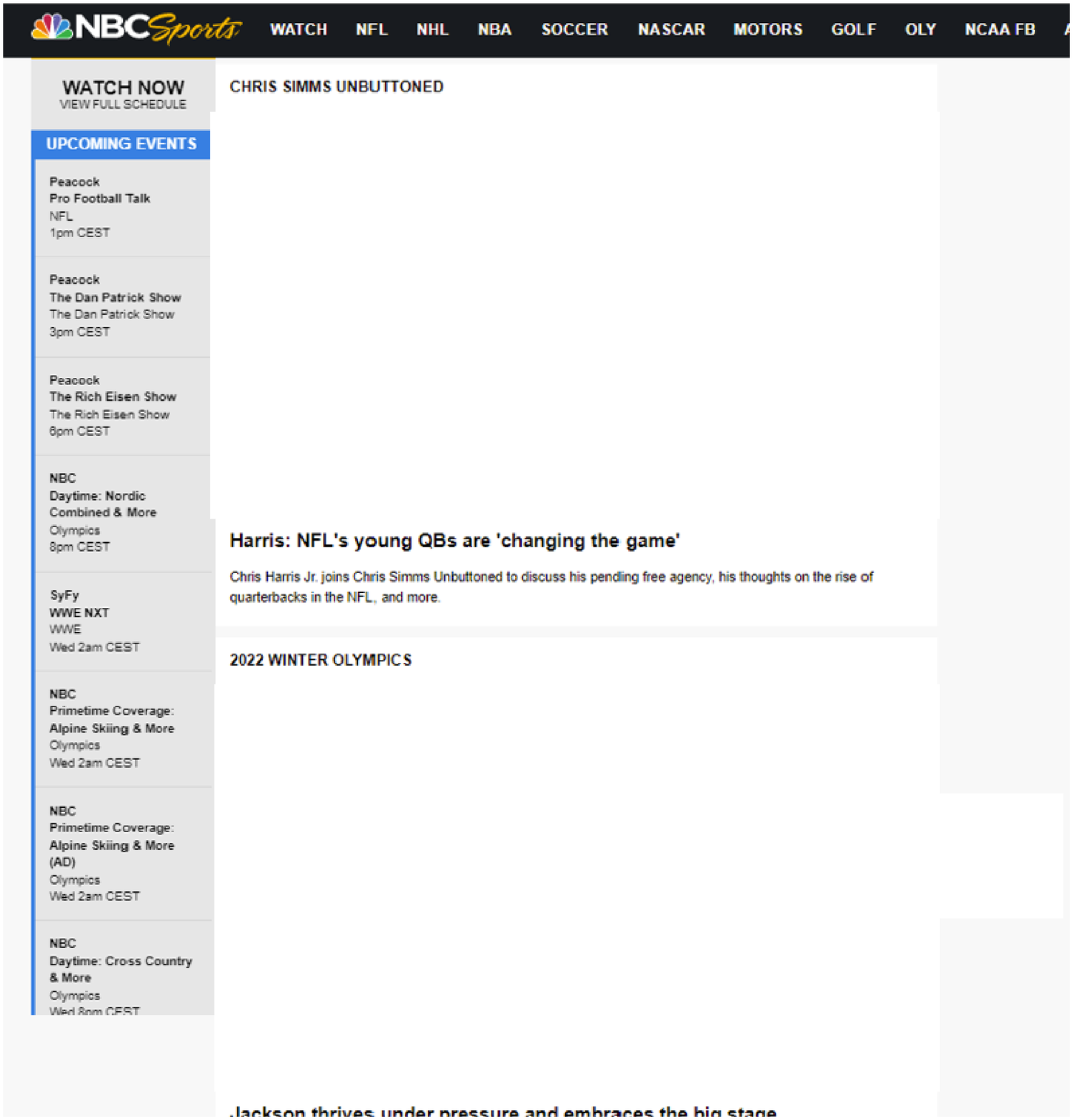}}\label{fig:sfig6}}
	\hfil
  \subfloat[Natural segments]
		{\frame{\includegraphics[height=3.5cm]{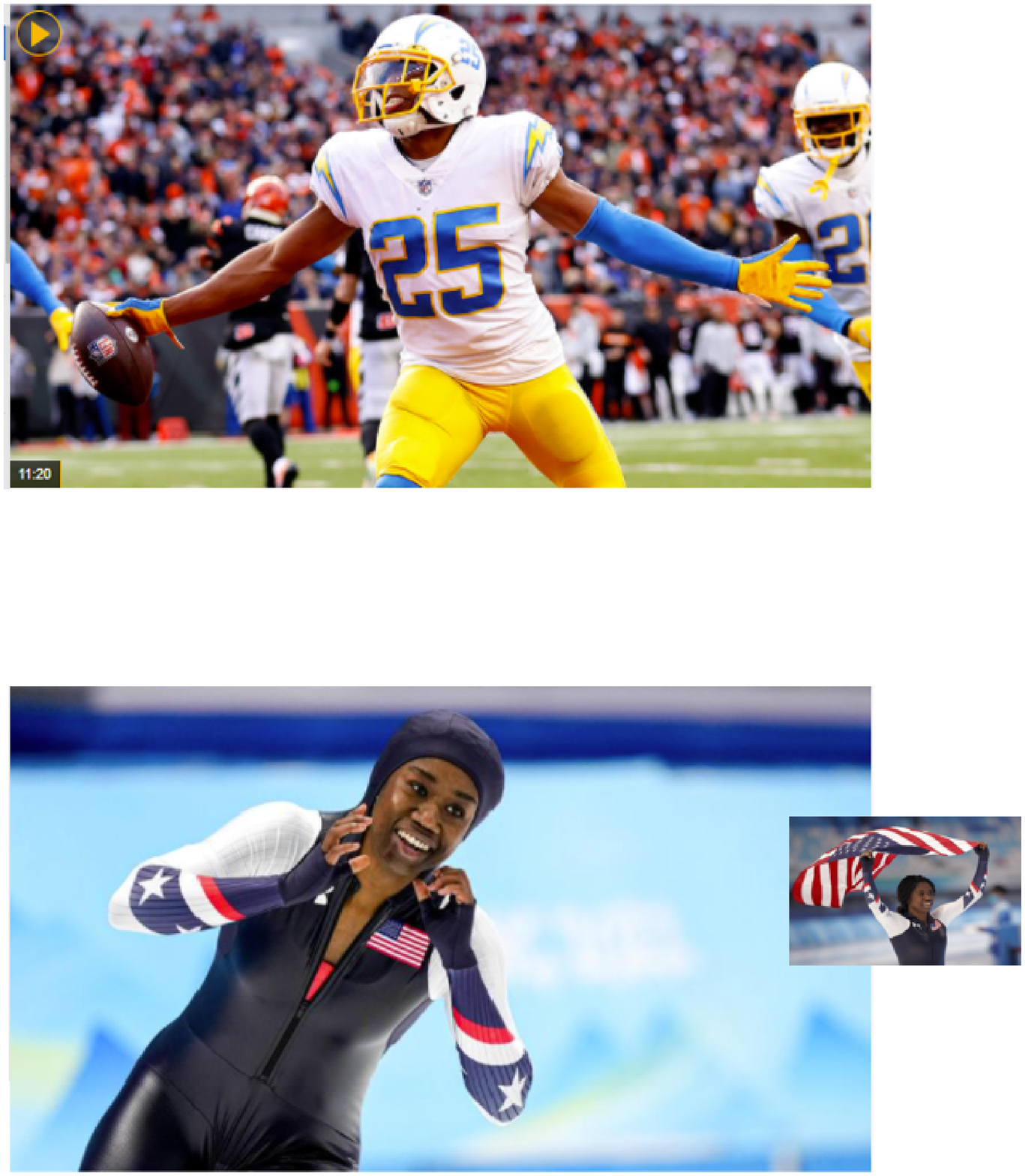}}\label{fig:sfig7}}
  \caption{Stages of segmentation (a-f) and processing (g, h)}
\label{fig_SegmentationSteps} ~\\[-1em]
\end{figure}

The block classification step splits the input image into non-overlapping blocks of $b \times b$ pixels, with $b = 16$, and classifies them as either natural ($> 128$ unique colours) or synthetic ($\le 128$ unique colours), \Figu{fig:sfig2}. 
A labelling function identifies objects of connected natural blocks (8-neighbourhood) and the corresponding bounding box information is derived, see \Figu{fig:sfig3}.
Such a bounding box must comprise at least 16 blocks, i.\ e.\ $b^2 \times 16 = 4096$ pixels. Otherwise, the segment is ignored in the following process, refer to {Figures \ref{fig:sfig3}} and \ref{fig:sfig4}. 

Furthermore, it is checked whether the percentage of natural blocks in the bounding box region is at least 60\% and if the average number of colours per block is at least 128. This filters out spurious segments, which just pass the initial thresholds without being real natural segments. This happens, for instance, if synthetic regions have artefacts or noisy textures. Now, all remaining segments undergo a process of refinement. This comprises the optimization of bounding box borders and the removal of overlaps, see \Figu{fig:sfig5}. These refinement steps are detailed in the following subsections.

\subsection{Segmentation refinement}	
\subsubsection{Extending or shrinking of bounding boxes}	
\label{subsec_exetendOrShrink}

Since non-overlapping image blocks are used for the determination of the initial bounding boxes, these boxes do not match with the exact border of the natural region, see \Figu{fig:sfig5} and \Figu{fig:closeUp}. It is possible that the natural pixels over the rows (or columns) of the bounding box border are missed or they cover synthetic content.
In order to segment the natural content more accurately, the extending or shrinking of the rectangular regions over the borders is necessary. 
The maximum number of rows (or columns) that can be reasonably added or removed in each direction is $b - 1 = 15$ with respect to the initial block size.

The refinement process is based on the 10 top most frequent RGB triplets observed in the entire image. It is presumed that these colours belong to the synthetic background, which holds true for majority of the images. The bounding box is extended as long as the pixels of the new row (or column) do not contain the top 10 colours, which means they are most likely natural pixels. If the majority of the pixels within a row (or column) contain one of the top 10 colours, this row (or column) is considered synthetic and is removed. 
The stop condition for extension is reached, when majority of pixels within the new row or column contain synthetic pixels. And the stop condition for shrinking is reached, when the pixels within the row or column contain a high number of natural pixels. 
The optimized bounding boxes can be seen better in magnified \Figu{fig:closeUp} (top and bottom left images).

\subsubsection{Removal of overlap}	
\label{subsec_removeOvrelap}

\begin{figure}
	\centering\frame{\includegraphics[height=7cm, width=80mm]{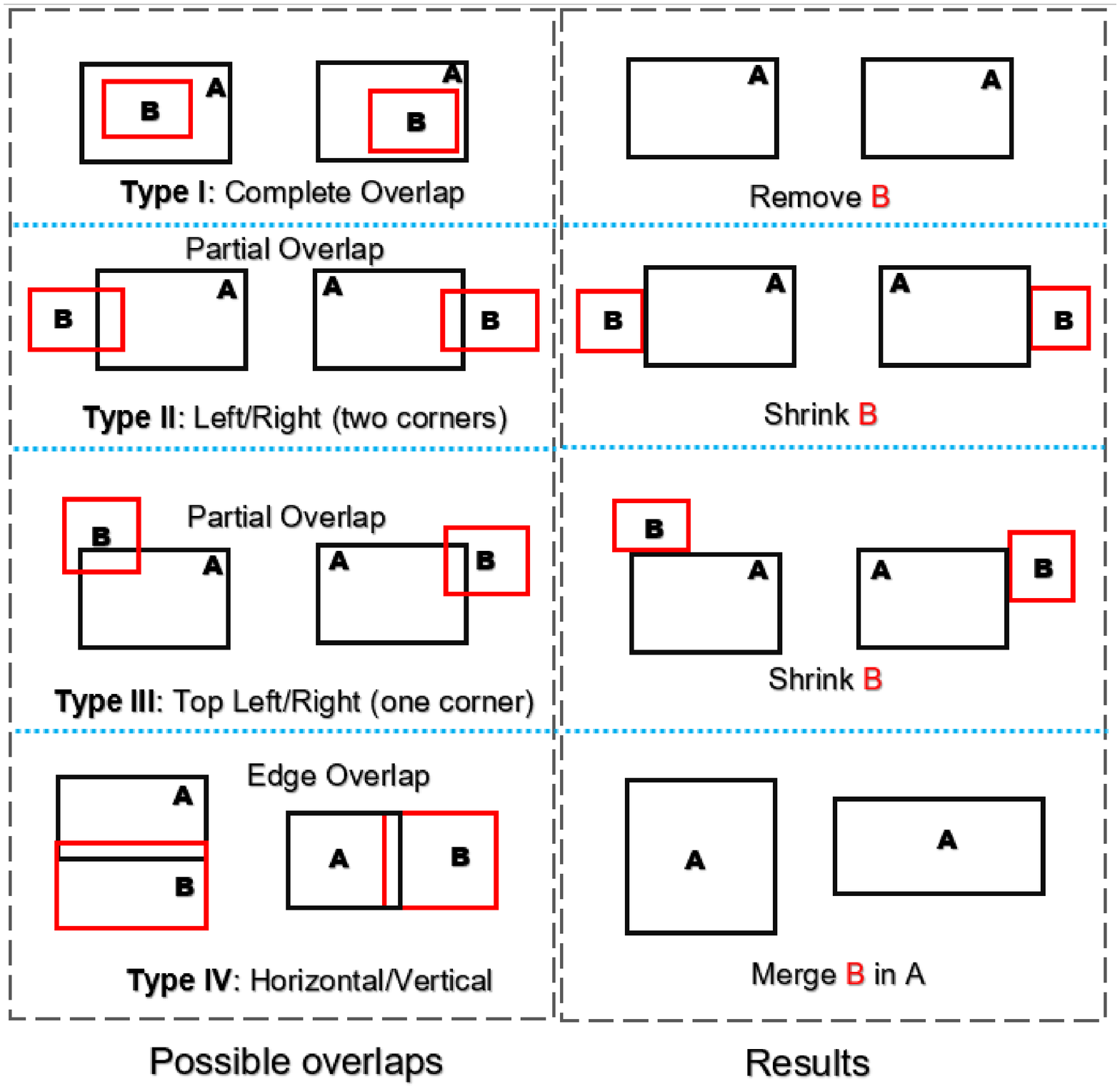}}
		\caption{Possible overlaps (left), Overlap removal (right)} 
		\label{fig:overlapping} ~\\[-1em]
\end{figure}
In \Figu{fig:sfig4}, it is seen that there are two overlapping rectangular regions at bottom half of the image. 
This can occur due to the presence of weakly connected natural blocks. In general, four types of overlap are possible, see \Figu{fig:overlapping}. The proposed method always retains the larger rectangular region and removes the smaller one (if a complete overlap occurs), resizes the smaller segment (if one or two corner overlap occurs), or merges both segments (if edge overlap occurs). Removal, resizing, or merging of these regions avoids unnecessary coding and transmission of bounding-box coordinates. 
\Figu{fig:sfig5} shows the refined bounding boxes after border refinement and overlap removal (with respect to \Figu{fig:sfig4}).

\subsection{Encoding/Decoding}	
\label{subsec_encodeDecode}

This subsection describes the modification of the coding process with respect to previous version of SCF described in \cite{Och21}. 
If there is at least one natural segment found within the image, the decoder is made aware of the following: Firstly, the segmentation flag is signalled cheaply with 1 bit. Secondly, the bounding box coordinates and lastly, the top most colour is transmitted. Sending this information takes about 40 bits per bounding box.  

In general, there are two options for the separate coding of synthetic and natural regions. Either the algorithm switches between different probability models depending on which region the current pixel belongs to or these regions are processed sequentially.
We decided to use the latter variant since the corresponding implementation allows a simpler management of the probability distributions.
Therefore, the coding process now comprises of two steps. First, a background image only containing the synthetic regions (\Figu{fig:sfig6}) is coded. Secondly, the image containing only the natural regions is coded (\Figu{fig:sfig7}). After decoding these two images, the decoder copies the natural segments back into the synthetic background image (reconstruction).

Regarding the adaptive estimation of the probability distribution, we took two of parameters into account: the pattern list with associated colour counts and the histogram of the global colour palette.
For both, there are three possibilities after coding the synthetic background. Either (i) the histograms are kept unchanged and the coding of the natural segments starts with the accumulated information derived from coding the synthetic background, or (ii) the counts of colours are reset to zero while the information about the existence of colours is kept, or (iii) all information is completely removed and knowledge has to be newly collected from scratch.
In total, there are $3^2=9$ combinatorial possibilities.
It turned out that on average the best compression performance can be achieved if the pattern list information is completely kept and the global colour palette is removed.
This behaviour can be explained as follows. If the natural segments do not share any statistics with the synthetic background, then they also show different patterns, and the collected information in the pattern list does not interfere with the new patterns. 
However, if the segments also contain background content, for example, because the natural region is not rectangular, the existing pattern list information can be useful.
The global colour palette, in contrast, affects the Stage 2 coding; by removing the background colours, the probability modelling can generate more narrow distributions for the encoding of natural pixels leading to lower bit rates. 


\section{Evaluation}	
\label{sec_investigation}

In this section, the proposed segmentation algorithm integrated into the SCF codec is compared to the SCF version from \cite{Och21} 
and also to HEVC with the screen content coding tools enabled \cite{HEVCSCC}. 
We have investigated a collection of 306 screen content images. The proposed segmentation method identifies 150 images with synthetic background and at least one natural segment. The other images are either considered as fully synthetic or natural.
This data set is assembled by pictures taken from internet resources (\cite{Set4, SCID}), different HEVC test sequences, and own collections. The data set is online accessible \cite{SCF_segmentation}. 
\Tabl{tab_comparison1} contains the lossless compression results of HEVC, the previous SCF version \cite{Och21} 
and the proposed version. 
	%
\begin{table}
\caption{\label{tab_comparison1} Comparison of the compression performance of the proposed method (SCF) with  HEVC (HM-16.21+SCM-8.8) and the previous version of SCF. The table lists sums of compressed file sizes in bytes.}
\centering
\begin{tabular}{|r|c|r|r|r|}
	\hline
Number\			&	HM-16.21		& \multicolumn{2}{|c|}{SCF}      \\
of images		&	SCM-8.8			& Previous \cite{Och21}  & Proposed             \\
\hline
150					& 42912198 		& 39055397 							 &  \bf{38470760}      \\
            & 111.6\%		& 101.52\%							 &  \bf{100.00\%}        \\
	\hline
\end{tabular} ~\\[-1.5em]  
\end{table}
	%


HEVC-SCC and VCC contain similar coding tools for lossless compression of screen content images and show a very similar compression performance in this context. Here, we achieve bitrate savings of 11.6\% and 1.52\% compared to HEVC and the previous version, respectively. The computational efforts for segmentation is vanishingly small compared to rest of the compression method.
\section{Conclusion and Outlook}	
\label{sec_conclusion}

Segmentation of natural content from the synthetic background and coding both regions separately helps in better estimation of probability distributions with respect to individual pixels. The investigations showcased in \Tabl{tab_comparison1} can be seen as a proof of concept that the segmentation approach can help to achieve better compression of SCIs. It has to be mentioned that the current version only detects and segments natural content from images with synthetic background and the classification here is only binary. In future work, we would like to extend our approach to segment synthetic regions from natural background images. In addition, at least a third class should be considered representing rendered images that are computer generated but still contain a very high number of colours. Using more than two classes has already been proposed in \cite{Din06}.
Furthermore, the encoding/decoding is jointly performed for all segments during the second round of processing. In cases where the segments have very different features, their separate treatment could be beneficial by different means. We also would like to extend the segmentation algorithm to deal with other shapes than rectangles. Processing natural segments using other lossless codecs which are dedicated to natural content could improve the compression efficiency of the entire system. 



\bibliographystyle{IEEEbib}
{
\bibliography{biblography}
}

\end{document}